\documentclass[12pt]{article}

\usepackage{graphicx}
\usepackage{amsfonts, amsmath}
\usepackage{setspace}
\newcommand{\be}{\begin{equation}} 
\newcommand{\ee}{\end{equation}}
\newcommand{\quadsp}{\;\;\;\;}
\newcommand{\boldt}{{\bf T}}
\newcommand{\Dslash}{\ensuremath \raisebox{0.08cm}{\slash}
  \hspace{-0.28cm}\nabla} 
\newcommand{\nslash}{\ensuremath \raisebox{-0.01cm}{\slash}
  \hspace{-0.21cm} n}
\newcommand{\goesto}{\rightarrow}

\begin{document}
\begin{titlepage}

%\today

%\rightline{HIP-2004-??/TH} 
\rightline{hep-th/0403217}

\vskip 3cm
\centerline{ \Large \bf Tachyon effective dynamics and de Sitter vacua}

\vskip 2cm

\centerline{ \normalsize Bruno Carneiro da
Cunha\footnote{bruno.cunha@helsinki.fi}}
\vskip .5cm
\centerline{\sl Helsinki Institute of Physics}
\centerline{\sl P.O. Box 64, FIN-00014 University of Helsinki, Finland}

\vskip 2cm

\begin{abstract}
We show that the DBI action for the singlet sector of the tachyon 
in two-dimensional string theory has a SL(2,R) symmetry, a real-time
counterpart of the ground ring. The action can be rewritten as that of
point particles moving in a de Sitter space, whose coordinates are
given by the value of the eigenvalue and time. The symmetry then
manifests as the isometry group of de Sitter space in two
dimensions. We use this fact to write the collective field theory for
a large number of branes, which has a natural interpretation as a
fermion field in this de Sitter space. After spending some time
building geometrical insight on facts about the condensation process,
the state corresponding to a sD-brane is identified and standard
results in quantum field theory in curved space-time are used to
compute the backreaction of the thermal background.
\end{abstract}

\end{titlepage}

\newpage

\setcounter{footnote}{0}

\section{D-brane tachyon dynamics}

The study of D-branes dynamics have shown that closed string degrees
of freedom can actually be hidden in the action of the open string
field theory. The recent re-interpretation of the c=1 matrix model as
an effective model of decay of D0-branes in two dimensional string
theory brought the novelty of studying non-stationary backgrounds in
which this duality is manifest, whose features we review as follows. 

The central point in the effective description is the DBI action:
\be
{\cal L} = -\int d^{p+1}x V(T)\sqrt{-\det G}
\label{dbi},
\ee
where the matrix $G$ has entries given by
\be
G_{ab}=\eta_{ab}+\nabla_a T \nabla_b T +\nabla_a Y^I\nabla_b
Y^I+F_{ab},
\ee
and $T, Y^I$ and $F_{ab}=\nabla_{[a} A_{b]}$ refer to the tachyon, the
compactification scalars and the gauge field living in the brane. The
Lagrangian (\ref{dbi}) was argued in \cite{Garousi:2000tr,
  Bergshoeff:2000dq, Kluson:2000iy, Sen:2002in, Sen:2002an,
  Sen:2002nu} to correctly describe the near on-shell behavior of the
tachyon condensation process.

The most striking fact about this model is that the
classical solutions of the potential:
\be
V(T) = \frac{1}{g_sl_s} \frac{1}{\cosh(\ell^{-1} T)}
\label{potential}, 
\ee
are solutions of the fully coupled string field theory
equations\footnote{\footnotesize provided $\ell=2 l_s$ in the
  bosonic string and  $\ell=\sqrt{2} l_s$ in the superstring.} 
in the approximation of homogeneous fields. This is so despite the
fact that at late times one would expect the field theory to be
strongly coupled and thus the classical picture not to hold. In a precise
sense (\ref{dbi}) is the unique Lagrangian which is both analytical in the
fields and gives rise to the correct classical equations of motion. For
further discussions see \cite{Kutasov:2003er} and references therein.
 
According to Sen's conjecture, the closed string excitations are hidden
in the open string dynamics. To recover it one starts with the
effective theory for the tachyon field of large number of branes and
study excitations for large values of the tachyon field. For
definiteness, consider the system of {\it N} D branes in non-critical
one-dimensional string theory, described by a non-abelian
generalization of (\ref{dbi}), a DBI matrix model quantum mechanical
system: 
\be
S[\{\boldt\}]=-\int dt~
\mbox{Tr}\left({\bf V}(\boldt)\sqrt{1-(D_t\boldt)^2}\right).\label{dbimm}
\ee
with $V(T)$ given as (\ref{potential}). McGreevy and Verlinde
\cite{McGreevy:2003kb} successfully reinterpreted the system above in
terms of a Liouville theory, in the limit where the Fermi sea level is
close to the top of the potential. In that limit, the system can be
approximated by the usual inverted harmonic oscillator model, where Sen's
conjecture help shed light on some old puzzles in the relation between
that model and two dimensional closed strings. In particular, an eigenvalue 
trickling down the maximum of $V(T)$, $T=0$ would represent a D0 brane
decaying to the closed string minimum. There is by now a great deal
of evidence supporting this point of view, in which the action above
can be recast in the double-scaling limit to an effective
two-dimensional theory in which the variables are closed string 
excitations.

In this approximation the model (\ref{dbimm}) indeed shows
perturbatively most of the 
behavior expected from the decay of the tachyon, gathered from
world-sheet techniques, like the production
of closed strings \cite{Lambert:2003zr}, the final state of the decay
process \cite{Klebanov:2003km}. Some non-perturbative effects where
also tied to the boundary states of Liouville proposed by Dorn-Otto
and Zamolodchikov-Zamolodchikov (DOZZ)
\cite{Martinec:2003ka, Alexandrov:2003nn}. As the description is
blurred in the strongly coupled regime, $T\approx \log g_s$, it made
sense to study (\ref{dbimm}) close to the open string vacuum
regime. One can then use
the incredible amount of data that has been gathered on this
particular model and make new interpretations in the light of the
McGreevy-Verlinde proposal. There is however a great deal of
information to be learned by considering generic Fermi sea profiles
\cite{Karczmarek:2003pv}, where one should deal with the fully coupled
problem directly. 

Heuristically, one would not naively expect (\ref{dbi}) to
hold. Backreaction from the closed string production should give large
contributions to the action and furthermore, in higher dimensions the
tachyon decay is non-homogeneous, a fact which is not naively
anticipated by the action above. Focussing for now on the first fact,
it is however surprising that the classical stress energy from
(\ref{dbi}) correctly describes the energy of the remnants of the
condensation process. And, since loops of (\ref{dbimm}) will involve
closed strings degrees of freedom in it, one is tempted to investigate
quantum effects of the action (\ref{dbimm}) starting from the point
where the open strings are treated classically. Working in two
dimensions will also help us in that one is then able to avoid the
problem of non-homogeneity. Also, building a direct correspondence 
between the effective open and closed string theories is also an
interesting problem in itself, and which should lend to a
amenable treatment in two dimensions.

The wishful thinking starts with considerations about the tachyon
effective dynamics. In this paper we start by considering a SL(2,R)
symmetry which allows us to write a collective field theory for the
tachyon excitations which is related to a field reparametrization to
that of \cite{McGreevy:2003kb}. The field has a classical
interpretation of eigenvalues moving in a conformally flat background,
with the potential playing the role of the conformal weight. In
Section 3 we consider generical reasons as to why this model is
not much different to the usual inverted harmonic oscillator as far as 
bosonization is concerned. In Section 4 we study finite energy
``probe'' solutions and the relations to the known facts about the
condensation process in two dimensions. In Section 5 we move on to
SL(2,R) invariant states and find a correspondence between the
Euclidean (Bunch-Davies) vacuum and the sD-brane state. We use
standard methods to compute the backreaction of this vacuum. We close
with speculative remarks about the fact that closed string excitations
arise as a ``near-horizon'' limit of the open string effective metric. 

As this manuscript was in its final stages of preparation we learned
of \cite{Berenstein:2004kk}, whose conclusions have similarities with
the points raised in Section 3 and 5. 

\section{Collective field theory of the DBI action}

A simple fact of the DBI action which has not been exploited so far in
the studies of tachyon condensation is that it displays an analogue of
the SU(2) symmetry present in the boundary state formalism
\cite{Callan:1994ub}. The symmetry appears as SL(2,R) which is more
suited to the real-time calculations we will be doing. At the quantum
mechanical level, the symmetry turns into a large symmetry algebra
which turns the model exactly solvable, much like in the
quintessential inverted harmonic oscillator case (see for instance
\cite{Jevicki:1993qn}). We will now explicit this algebra for (\ref{dbimm}).

The model in (\ref{dbimm}) has an underlying $U(N)$ symmetry that is
gauged by a non-dynamical field (hence the covariant derivative in
(\ref{dbimm})). If we pick the Coulomb gauge for this field the term
turns into a Lagrange multiplier that projects the field $T$ into
singlet states, i.e., the dynamics then depends only on the set of
eigenvalues of $\boldt$, $\{\tau_i\}$, see
  \cite{Klebanov:1991qa,Polchinski:1994mb} for a thorough
  review. Since only the eigenvalues are dynamic, $\dot{\boldt}$ and
  $\boldt$ commute and we can rewrite the action as:
\be
{\cal L}_{\rm cl} = - \sum_{k=1}^N V(\tau_k)\sqrt{1-\dot{\tau}_k^2}
\label{classicalaction},
\ee
for a generic potential ${\bf V}(\boldt)$ which commutes with
$\boldt$. It is also important
to point out that in the path integral formalism, the change of variables
from the matrix elements to the eigenvalues introduces a Vandermonde
determinant which enforces Fermi statistics in the wave
functions. Absorbing this determinant into the measure leads us to an
anti-commuting field of eigenvalues. One
could use the collective field formalism and re-express the
action in terms of a density field 
\be
\rho(\tau,t) = \int dk e^{ikx}\mbox{Tr}(e^{-ik \boldt(t)})=\sum_{k=1}^N
\delta(x-\tau_k(t)),
\ee
but we will be a little more heuristic. One needs to Schr\"odinger
equation, with a Hamiltonian for a single eigenvalue given by:
\be
\hat{H}(\pi,\tau) = \sqrt{{\hat{\pi}^2+V(\tau)^2}} \label{singlefermion}.
\ee
But the definition of (\ref{singlefermion}) as an operator suffers
from ordering ambiguities. We are now facing the same
problem as Dirac: how to take the square root of an operator while
keeping the Pauli exclusion principle. This is now complicated
by the fact that $\pi$ and $V(\tau)$ neither (anti-)commute, nor are
divisors of zero in the operator algebra.

In order to solve this problem we note that the action
(\ref{classicalaction}) with the potential (\ref{potential}) has an
SL(2,R) symmetry. In fact it describes classical motion of particles
of mass $m=1/g_sl_s$ in a curved background:
\be
S_{\rm cl}=-\sum_k\int dt_k~
\frac{1}{g_sl_s}\sqrt{\frac{1}{\cosh^2(\ell^{-1}\tau_k)}-
\frac{\dot{\tau}_k^2}{\cosh^2(\ell^{-1}\tau_k)}}\equiv -\sum_k
\int dt_k~ m \sqrt{\left(\frac{ds_k}{dt}\right)^2}
\label{cldsp},
\ee
where the metric, given by:
\be
ds^2=\frac{1}{\cosh^2(\ell^{-1} \tau)}(dt^2-d\tau^2) \label{metric},
\ee
can be mapped to the static patch of a two-dimensional
de Sitter space. Defining $r=\ell \tanh(\ell^{-1} \tau)$:
\be
ds^2=(1-(\ell^{-1} r)^2)dt^2-\frac{dr^2}{1-(\ell^{-1} r)^2}
\label{static}, 
\ee
where $-\ell < r< \ell$ maps to the whole real line in the $\tau$
coordinate. The SL(2,R) algebra can be seen as isometries of the
``space-time'' whose coordinates are given by time $t$ and the eigenvalue
$\tau$. Thus any classical trajectory can be obtained by acting
with the SL(2,R) generators on the static trajectory, perched at the
top of the potential $\tau=0$. 

To be more explicit, we can forget for a moment the Fermi statistics
and the ambiguities of defining a quantum Hamiltonian and numbly square  
(\ref{singlefermion}) with the potential (\ref{potential}). The resulting
equation of motion can be written as:
\be 
\left(\cosh^2(\ell^{-1}
\tau)[\partial_t^2-\partial_\tau^2]-m^2\right)
\phi(\tau,t)=0  
\label{klein-gordon},
\ee
where we call the double-scaled mass also $m$, by an abuse of
language. Now, by introducing the operators:
\be
J_\pm  =  \ell e^{\pm \ell^{-1} t}\left(\cosh(\ell^{-1}
\tau)\frac{\partial}{\partial \tau}\pm
\sinh(\ell^{-1} \tau)\frac{\partial}{\partial t}\right),\quadsp
J_3 = \ell \frac{\partial}{\partial t} \label{generators},
\ee
the equation (\ref{klein-gordon}) turns into the suggestive form of:
\be
{\bf J}^2\psi(x)=\left(\frac{1}{2}(J_+ J_- + J_- J_+)
-J_3^2\right)\phi(x)= m^2\ell^2\phi(x).
\ee
The operators in (\ref{generators}) do
indeed satisfy a SL(2,R) algebra: 
\be
[J_3, J_\pm]=\pm J_\pm,\quadsp [J_+, J_-]=2J_3.
\ee
With these provisions, the natural candidate for the quantized
collective field action will be the free, massive Majorana fermionic
field in the cosmological patch of two dimensional de Sitter space. The
action will then be:
\be
S_{\psi}=\frac{1}{2}\int d^2x~
e~\left(i\bar{\psi}{\bf \gamma}^a\nabla_a\psi
-i(\nabla_a\bar{\psi}){\bf \gamma}^a\psi - 2 m
\bar{\psi}\psi\right) \label{effaction},
\ee
where $e=\det[e^i_a]$ is the determinant of the zweibein,
${\bf \gamma}^a=e^a_i{\bf \gamma}^i$ are the de Sitter gamma
matrices\footnote{They are taken to satisfy
  $\{{\bf \gamma}^a,{\bf \gamma}^b\}=2g^{ab}$. See the Appendix for
  conventions and a review of the needed spinor calculus.} and 
\be
\nabla_a\psi=\partial_a\psi-\frac{1}{2}
\Sigma^{ij}e^b_i(\nabla_a e_{bj})\psi
\ee
is the covariant derivative compatible with the metric
(\ref{metric}), with $\Sigma^{ij}$ being the Lorentz generator. As
a matter of fact the action (\ref{effaction}) is very similar to the
action proposed in \cite{McGreevy:2003kb}:
\be
S_{\chi}=\int d^2x~\left(\bar{\chi}\gamma^i\partial_i\chi-
\frac{m}{\cosh\ell^{-1}\tau}\bar{\chi}\chi\right),\label{mcg-vact}
\ee
except that in (\ref{effaction}) the SL(2,R) symmetry of the classical
solutions is manifest. In fact one recovers \cite{McGreevy:2003kb}  by 
rewriting the action above with the metric (\ref{metric}) in terms of
the scale invariant fermionic field:
\be
\chi(t,\tau) = \cosh^{-1/2}(\ell^{-1}\tau)\psi(t,\tau).
\ee
One recovers (\ref{mcg-vact}) since in two dimensions the
action $S_\psi$ does not depend on the spin connection. Since we are
interested in the symmetry itself, we will work with (\ref{effaction})
from now on. 

The algebra given by
(\ref{generators}) is augmented by addition of the $w_{\infty}$ generators:
\be
{\cal O}_{n,m}=J_+^{n+m}J_-^{n-m},
\ee
with $n=1,\ldots$ and $-n<m<n$. The structure hints at an underlying
Lax formalism and then at the integrability of the theory, which would
be hardly surprising since the recast theory is equivalent to free
massive fermions. The occurrence to $w_\infty$ algebras is usually
associated to the presence of massless scalars. One has to wonder then
if there is an alternative way of recasting the Lagrangian
(\ref{effaction}) in terms of a massless field, just like in the case
of massive Klein-Gordon field in flat two-dimensional space
\cite{Ginsparg:is}. Clarifying this point would be interesting to the
problem of bosonization.

As a preamble for the discussion in the next section, one is
interested in the regime where the observer at $\tau=0$ sees no
excitations, because classically none of them can climb up the
potential wall. The instability of the open string vacuum correspond
to the fact that in de Sitter near light-like geodesics repel and thus
any wavefunction that is initially perched at the top of the potential
will at the end ``spread over'' and run off to the horizon at
$r=\pm\ell$, or $\tau=\pm\infty$. The latter is identified with 
the closed string vacuum.

There is a curious aspect of two dimensional de Sitter spaces which
hinders the study the system in terms of the proper-time of the
eigenvalues as they approach the horizon. The reparametrization
$\tau\goesto r$. shows rather clearly that the point $\tau=\infty$ is
at finite distance in field space, but also does not say anything
about what happens after that. The question arises as what is the
metric beyond the horizon. The answer since there is a family of
SL(2,R) invariant spaces that allow for the same static
patch\footnote{I thank Will McElgin for this remark.}. For instance, 
all spaces considered in \cite{DaCunha:2003fm} will admit a static
patch of the sort described above (\ref{static}) if they are of the
elliptic class and $\epsilon>\frac{1}{2}$ in
\be
ds^2=\frac{-dt^2+d\sigma^2}{\sin^2 \epsilon t}.\label{cylinder}
\ee
Thus one cannot use the classical SL(2,R) invariance to predict what
will happen at the end of the decay process. We will however use
quantum mechanics to infer about the global structure of field space
in Section 5.

\section{A few classical remarks}

Before continuing to explore the symmetry of the DBI action, we should
stop for a moment and try to understand the relation between this
description of the tachyon dynamics and the usual inverted harmonic
oscillator model. There one considers a number of non-relativistic
fermions running along with the Hamiltonian:
\be
H(p,q)=\frac{p^2}{2}-\frac{q^2}{2}+\frac{1}{g_s}\label{invho},
\ee
where we fixed the additive constant to have $H=0$ at the Fermi
surface. As discussed in \cite{Sen:strings2003}, there is a canonical
transformation relating (\ref{invho}) to (\ref{cldsp}). In order to
obtain it one just needs to solve each system separately as a function
of energy and initial time:
\be
\begin{array}{l}
p=p(E,t) \\
q=q(E,t)
\end{array}
\Leftrightarrow
\begin{array}{l}
\pi=\pi(E,t) \\
\tau=\tau(E,t)
\end{array}.
\ee
One can then eliminate $E$ and $t$ and find the transformation
$p(\pi,\tau)$, $q(\pi,\tau)$. This transformation
does not map the solutions of (\ref{invho}) with $H<0$, but they show
up again in the relativistic model when we consider the quantum
mechanical effective action (\ref{effaction}), which can be seen as an
advantage of this formulation. On the other hand, the Fermi level
$H=0$ is mapped to $\tau\goesto \infty$ in the relativistic
Hamiltonian and then it may be tricky to extract information about the
closed string vacuum in this case. However, in (\ref{cldsp}) the
closed string vacuum corresponds to the horizon,
as one can see from (\ref{static}) taking into account that
$\tau\goesto\infty$ maps to $r=\ell$. So the transformation
above really maps small excitations around the closed string vacuum,
like for instance the dilaton or the ``tachyon'',  to ``near horizon''
excitations in the relativistic model. Extracting information about
closed strings excitations from (\ref{cldsp}) should be equivalent to
studying excitations of the field (\ref{effaction}) around the horizon
of de Sitter, a well-known problem in QFT in curved space-time. 

The argument can be made more explicit by actually computing
$q(\pi,\tau)$:
\be
q=\sqrt{2\left(\frac{1}{g_sl_s}-E\right)}\cosh\left[\ell
\;\mbox{arccosh}\left(\frac{Eg_sl_s}{\sqrt{1-E^2g_s^2l_s^2}}
  \sinh\frac{\tau}{\ell}\right)\right].
\ee
At large values of the position, $\tau$ becomes the ``time of flight''
coordinate of (\ref{invho}) \cite{Klebanov:1991qa}, so in the ``near
horizon'' limit there is a clear connection between the excitations of
the relativistic fermions of (\ref{cldsp}) and the closed string
fields. Here we used the time translation symmetry of the two systems
to hide constants. In hindsight this is not so unexpected since the
potential drops to zero exponentially and hence one should be able to
describe the system as free massless fermions as
$\tau\goesto\pm\infty$, in which case would presumably know how to
bosonize the system. One notes that the transformation $(p,q)\goesto
(\pi,\tau)$ is defined only for positive energy solutions. How exactly
one should include negative energy solutions and the other side of the
potential $q<0$ will rely on extra input, like a comparison between
reflection coefficients of the quantum mechanical model, which will be
postponed until the next section. See also \cite{Das:1990ka} for the
relation between the collective field theory of (\ref{invho}) and the
relativistic field theory (in flat space).

Nevertheless this does not mean that one is required to take the
``near horizon'' limit to talk about closed string excitations. We
begin to illustrate this point by using the classical integrability of
the system (\ref{cldsp}). The Hamilton-Jacobi equation can be solved
to give
\be
S_{\rm cl}(t,\tau)=-Et+\int_{\tau_0}^\tau
d\tau'\;\sqrt{E^2-V(\tau')^2}
\label{clasvalueaction},
\ee
which can be written as a sum of elliptic integrals. The classical
action (\ref{clasvalueaction}) generates the canonical transformation
that trivializes the dynamics. Any such system can be mapped to
(\ref{invho}) by composing the transformations, as long as the spectra
of energies match. Then one can rewrite the ``loop functional'':
\be
W(\ell, t)=\int_R dp\wedge dq \; e^{\ell q},
\ee
integrated over the volume of phase space occupied by the Fermi
sea. The integral above, originally written in terms of the $(p,q)$
variables of (\ref{invho}), can thus be recast in terms of any
integrable system since the canonical transformation preserves the
volume of phase space and the function $q$ can be written in the new
set of coordinates, (\ref{cldsp}) being but one case among an infinite
number. If the Fermi level is at a surface of constant energy $H=\omega$,
variations of $W$ satisfy the Wheeler-de Witt equation
\cite{Ginsparg:is}:
\be
\left(\partial^2_t-\partial^2_\ell\right)\delta
W=2\omega\ell^2\;\delta W ,
\ee
which can be written without any reference to the parametrization of
the matrix model. The Liouville field can be obtained from $\delta W$
by the usual procedure \cite{Moore:1991ag}, at least in the genus
zero, or zero string coupling approximation. This trick will of course
not have a natural interpretation in terms of density of eigenvalues,
or loop observables of the matrix model, but serves as a reminder that
one can reshuffle the degrees of freedom in various ways.

One can then see the effective action for fermions
(\ref{classicalaction}) as a model which interpolates between the
inverted harmonic oscillator model and the relativistic fermions of
\cite{Das:1990ka}. The model is also equivalent classically to the
inverted harmonic oscillator, related to it by a finite canonical
transformation, generated by the ground ring of (\ref{invho})
\cite{Witten:1991zd, Douglas:2003up}. The non-trivial value of the
observation weights on the fact that this model also has a ground ring
(\ref{generators}), realized here as generators of a ``space-time''
symmetry. One should bear in mind that this non-flat background is not
the one on which the closed degrees of freedom themselves exist, like
the one proposed in  \cite{Strominger:2003tm}, but rather an effective
way of looking at the dynamics of the open strings. It should be
pointed out, however, that one could in principle obtain a adS$_2$
background in a similar fashion. The lesson from the BCFT studies is
that the tachyon potential gives rise to a constant curvature in field
space. By considering space-like tachyon condensation, this curvature
will become negative, and thus the relevant geometry of the field
space will be as in the paper cited above. In other words, the adS$_2$
metric can be obtained from (\ref{metric}) by a double Wick
rotation. This seems to be an interesting new route to explore in
further research.

We close this section with a remark on the spectrum of the action
(\ref{effaction}). Although the theory is relativistic and
massive, there are excitations with arbitrarily low energy in
it. These can be understood classically. Consider the classical
trajectory $A$ in Fig. 1. In the usual tachyon dynamics language it
represents a D0 brane perched at the top of the potential. Its energy
is just $m$. In the de Sitter interpretation it just corresponds
to a particle at rest at the ``observer's position'', $\tau=0$. 

\begin{figure}[hbt]
\begin{center}
\mbox{
\includegraphics[height=6cm]{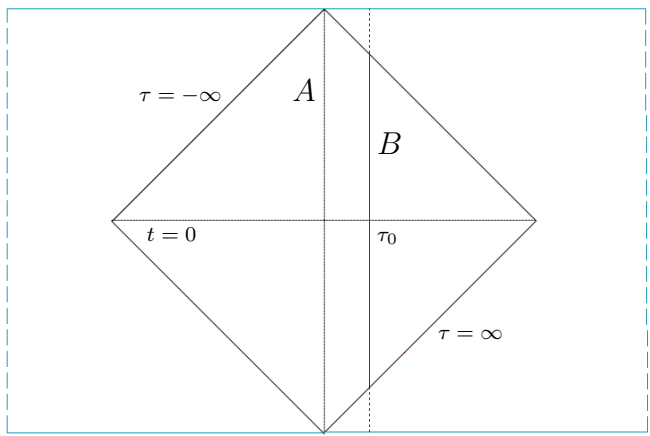}
}
\put(-145,130){$A$}
\put(-113,110){$B$}
\put(-200,77){\scriptsize $t=0$}
\put(-203,130){\scriptsize $\tau=-\infty$} 
\put(-90,40){\scriptsize $\tau=\infty$}
\put(-113,77){\scriptsize $\tau_0$}

\caption{ \it The conformal diagram of 2-dimensional de Sitter. The
  diamond represents the static patch, covered by the
  coordinates $t,\tau$. Two time-like geodesics are drawn, $A$ is
  perched at $\tau=0$, whereas the ``bounce'' $B$ has non-trivial $\tau$
  dependence. The $B$ is obtained from $A$ by a  suitable
  application of the generators (\ref{generators}).}
\end{center}
\end{figure}

Consider now the ``bounce'' trajectory, $B$. It represents an
eigenvalue coming from $\tau=\infty$ and scattering off of the
tachyon wall, a classical bounce trajectory. It can also interpreted
as a particle ``at rest'', {\it   i.e.}, following another geodesic in
de Sitter. A local observer at any point of $B$ will see it as having
energy $m$, but an observer at $A$ ($\tau=0$) will see its energy {\em
  redshifted} according to the non-triviality of the metric:
\be
E=\frac{m}{\cosh{\ell^{-1}\tau_0}}\label{energy},
\ee
where $\tau_0$ is the turning point of the classical
trajectory. Then, by ``pushing away'' geodesics to the horizon one can
make excitations of arbitrarily low energies as measured by $A$. The
generators that shift the trajectories are none other than $J_\pm$ in
(\ref{generators}).

\section{Finite energy solutions}

In this section we will make the relation between (\ref{effaction})
and (\ref{mcg-vact} precise, and illustrate the
integrable structure. The relation between the Dirac equation in two
dimensions and integrable models is of course not new
\cite{Cooper:az}, and it applies to any functional form of the DBI
potential. The relation between shape-invariant potentials and the
condensation process has also been studied in
\cite{Zwiebach:2000dk}. The exercise will also provide the scattering
amplitudes which are usually used to compute expectation values of the
closed string observables \cite{Moore:1991zv}. Consider then solutions
of  
\be
(i\gamma^a\nabla_a-m)\psi=0\label{equation},
\ee
with constant energy:
\be
{\cal L}_t\psi_\omega=-i\omega\psi_\omega\label{frequency}.
\ee
For this purpose it will be interesting to work in a conformally flat
metric, like (\ref{metric}). In the metric
$ds^2=\exp 2\rho(x)(dt^2-dx^2)$ the spin connection is given by
\be
{\bf \Gamma}_a=-\frac{1}{2}\rho'\gamma^5(dt)_a.
\ee
Multiplying (\ref{equation}) by the flat-space $\gamma^0$ we will have:
\be
(i\partial_t-\frac{i}{2}\rho'\gamma^5-i\gamma^5\partial_x-me^\rho\gamma^0)
\psi=0.
\ee
After writing
$\psi(x,t)=\exp(-1/2\rho(x))\exp(-i\omega t)\chi_\omega(x)$, one
arrives at an eigenvalue problem for $\chi_\omega$
\be
(i\gamma^5\partial_x+me^\rho\gamma^0)\chi_\omega(x)=\omega\chi_\omega(x).
\ee
Since we are working in two dimensions, and $\gamma^5$ anti-commutes
with $\gamma^0$, one can choose a basis of Dirac matrices where
$\gamma^5=\sigma^2$ and $\gamma^0=\sigma^1$. The system now reads
\be
\begin{array}{c}
Q\chi^+_\omega=\omega\chi^-_\omega \\
Q^\dagger\chi^-_\omega=\omega\chi^+_\omega 
\end{array},\label{susyqm}
\ee
where $Q=\frac{d}{dx}+m e^\rho$ and
$\chi_\omega=\chi^+_\omega\eta_++\chi^-_\omega\eta_-$ with $\eta^\pm$
a constant normalized spinor $\eta_\pm=\pm i\gamma^1\eta_\pm$ . Note
that the system has the ``S-like-duality'' \cite{Douglas:2003up},
$\chi^\pm(-x)=\chi^\mp(x)$. Thus charge conjugation of the fermionic
field, represented here by $\sigma^3$ \cite{Jackiw:1975fn}, is mapped
to dualization of the compact boson $C_0$ in the closed string
formulation.

For the particular form of the DBI potential (\ref{potential}),
$e^{-\rho(x)}=\cosh \ell^{-1}x$, the system above was solved using the
factorization method in  \cite{Khare:tg}\footnote{We follow
  conventions where $_2F_1(a,b;c;0)=1$ \cite{abramovitz}.}. One finds
two independent solutions for the equation $Q^\dagger
Q\chi^+_\omega=\omega^2\chi^+_\omega$, namely 
\begin{eqnarray*}
S_1(x)& = & e^{m\ell \arctan z}
  {_2F_1}(i\omega\ell,-i\omega\ell;{\textstyle
  \frac{1}{2}-im\ell;\frac{1}{2}(1+iz)}) \\ 
S_2(x)&\!\!\!\!\!\! = &\!\!\!\!\!\! e^{m\ell \arctan z}\!\! 
  \left(\frac{1+iz}{2}\right)^{\frac{1}{2}+i\omega\ell}\!\!\!\!{_2F_1}(
{\textstyle  \frac{1}{2}+i(m+\omega)\ell, \frac{1}{2}+i(m-\omega)\ell;
  \frac{3}{2}+im\ell; \frac{1}{2}(1+iz)}),
\end{eqnarray*}
where $z=\sinh\ell^{-1}x$. The asymptotic expansion of these solutions
is a sum of plane waves:
\be
\lim_{x\goesto\pm\infty}S_i(x)= a_{i\pm}e^{i\omega x}+b_{i\pm}e^{-i\omega x}
\ee
but we refer to \cite{Khare:tg} for the exact formulas for the
coefficients $a_{i\pm}$ and $b_{i\pm}$. We also quote the result for
the reflection coefficient: 
\be
R^+(\omega)=\frac{\sinh\pi m\ell}{\cosh \omega\ell}
\frac{\Gamma(\frac{1}{2}+i(m-\omega)\ell)
  \Gamma(\frac{1}{2}-i(m+\omega)\ell)}{
  \Gamma^2(\frac{1}{2}-i\omega\ell)} \label{reflection},
\ee
with $\chi^-_\omega$ having the opposite sign for the reflection
coefficient. From the decomposition of the spinor one sees the
incident and reflected wave have opposite chirality. In the usual
inverted harmonic oscillator model, this coefficient, along with the
SL(2,R) generators (\ref{generators}) compound in the limit
$m\rightarrow\infty$ ($g_s\rightarrow 0$),the integrable structure via
the scattering operator $\hat{S}$, defined in the usual way:
\be
\hat{S}\chi^+_{+\omega} = R(\omega)\chi^+_{-\omega}.
\ee

One can see directly from (\ref{reflection}) the strength of
non-perturbative effects, due to tunneling through the top of the
potential \cite{McGreevy:2003kb}. 
\be
|R(\omega)|^2=1-4e^{-2\pi m\ell}\cosh^2 \pi\omega\ell+{\cal
  O}(e^{-4\pi m\ell}).
\ee
In the de Sitter interpretation, these correspond to the flux of
particles coming from one horizon and tunneling to the antipodal
horizon without being detected at $\tau=0$. Furthermore, the phase of
the amplitude, again in the $m\goesto\infty$ ($g_s\goesto 0$) limit, 
\be
R^+(\omega)\approx
(m\ell)^{2i\omega\ell}\frac{\Gamma(\frac{1}{2}+ 
  i\omega\ell)}{\Gamma(\frac{1}{2}-i\omega\ell)}.
\ee
As expected, this is exactly the same expression
one finds in type 0B, apart from the piece of the phase linear in
$\omega$ \cite{Douglas:2003up}, giving rise to the same density of
states as a function of the energy. From the reflection coefficients
one can then compute scattering amplitudes of the fermion densities
\cite{Moore:1991zv} by combining chiral components. Perturbatively the
excitations at $\tau>0$ and at $\tau<0$ are also independent. As
alluded above, non-trivial profiles for the region $\tau<0$ are mapped
to non-trivial profiles of the RR-scalar in the closed string
formulation. This is natural since the transmitted wave has to tunnel
through the potential, thus describing a D-instanton which couples to
the scalar.

Despite being natural from the point of view of
\cite{McGreevy:2003kb}, the ``Rindler'' vacuum obtained from
considering the state with no positive frequency modes as measured by
the ``open string observer'' (\ref{frequency}) gives rise to an
infinite backreaction when quantum effects are computed. With the
interpretation given in the last section, this effect can be seen as
the backreaction in terms of closed strings. In fact, if can
extrapolate results for scalar fields, the renormalized stress energy
diverges at the horizon, where the closed string excitations are
supposed to be localized.

As a closing note to this section, one should point out that the
picture changes dramatically when one considers the case of the  
bosonic string. In the inverted harmonic oscillator model one does not
fill the Fermi sea at one side of the potential, with the result does
not have a bounded spectrum. Here there is no clear way to perform
an extra truncation of the spectrum, since the scattering process
mixes solutions with positive and negative chirality. Also, one sees
that the scattering amplitudes as found above would disagree with the
usual bosonic calculations. So it does not seem feasible to realize the 
bosonic string as the effective model of a large number of D particles
as we are exploring here. One would not find this entirely surprising,
as the argument of non-analyticity of the DBI action was raised in
\cite{Kutasov:2003er}, but it would be interesting to clarify the
discrepancies between those points of view.

\section{Reinterpreting the Euclidean vacuum}

In the studies of QFT in de Sitter background one pays particular
attention to de Sitter invariant states. Those are the analogue of the
``vacuum'' state in usual field theory, although there is no good
notion of ``no particle state'' generically. The physical content of
the SL(2,R) invariant states can be studied by means of the symmetric,
or Hadamard two-point function: \cite{birrell}: 
\be
\langle \alpha | \{\phi(p),\phi(p')\} | \alpha \rangle = G(p, p'),
\ee
which in the particular case of SL(2,R) invariant states will depend
only on the SL(2,R) invariant quantity, $Z$, given by:
\be
Z(\tau,t;\tau',t')=
\frac{\cosh\ell^{-1}(t-t')+\sinh\ell^{-1} \tau\sinh\ell^{-1}
  \tau'}{\cosh\ell^{-1}\tau\;\cosh\ell^{-1}\tau'} \label{zed} ,
\ee
when the metric of the manifold is taken to be (\ref{metric}). $Z$ is
related to the geodesic distance $\mu(p,p')$ between $p$ and $p'$ by
\cite{Allen:ux}: 
\be
 Z(p,p')=\cosh \frac{\mu(p,p')}{\ell}.
\ee
For instance, for time-like geodesics, $\mu$ is the proper time. In
this way we can find any solution to the classical equations of
motion, provided we also consider the direction of movement. To
illustrate this point, consider the geodesic whose turning point is
$p_0=(t_0,\tau_0)$. Then by solving the equation
$Z(p,p_0)=\cosh\ell^{-1}\mu$ one arrives at the geodesic:
\be
\sinh\frac{\tau-\tau_0}{\ell}\tanh\frac{\tau_0}{\ell} =
\cosh\frac{t-t_0}{\ell}.
\ee
The energy of this solution is just (\ref{energy}). The geodesic
distance $\mu$ provides a geometrical interpretation for the
uniformizing variable $s$ that shows up in matrix elements of loop
operators in \cite{McGreevy:2003ep}. As we will see in this section,
it will be very useful to re-express SL(2,R) invariant quantities as
functions of $Z$.

If we rewrite the Klein-Gordon equation (\ref{klein-gordon}) in terms
of $\mu$, we will have:
\be
\frac{1}{\sinh\frac{\mu}{\ell}}\frac{d}{d\mu}\left(\sinh\frac{\mu}{\ell}
\frac{d}{d\mu}G(\mu)\right)+m^2 G(\mu)=0, \label{laplacian}
\ee
which is satisfied by both Wightman functions (positive and
negative frequency two-point functions) as well as their sum, the
Hadamard function. Its solutions are conical functions:
\be
G(Z)=a P_{-\frac{1}{2}+is}(Z)+b P_{-\frac{1}{2}+is}(-Z)
\label{solution0energy},
\ee
with $s^2=(m\ell)^2-1/4$. When seen as a Hadamard function,
each solution (\ref{solution0energy}) defines an SL(2,R) invariant
state, all of which can be seen as a Bogoliubov transformation of the
Euclidean, $b=0$ vacuum. The solution for $b\neq 0$ does not have the
Hadamard form close to the light cone $\mu(p,p')=0$ and hence does
not have a sensible field theory interpretation as one approaches the
UV scale\footnote{In particular, one cannot generically define a
  sensible perturbation theory around a generic solution. See
  \cite{Einhorn:2002nu} for details.}. The Feynman function is
obtained by evaluating the function (\ref{solution0energy}) just above
the branch cut $Z\in [1,\infty[$, that is to say, to take
    $G(Z+i\epsilon)$. For details about the delta-function
    representation see \cite{Candelas:du}. 

There is a relation between the Euclidean two-point function, with
$b=0$, and the zero energy solution $J_+\phi = 0$:
\be
\phi_0(t,\tau)=P_{-\frac{1}{2}+is}(\tanh \ell^{-1}\tau).
\ee
One can obtain the equation above from (\ref{solution0energy}) by
means of an analytical continuation. When the separation of the two
points is space-like, $\mu^2(p,p')<0$, then one can consider the two
points to be at a equal-time hypersurface, so $t=t'$. The zero energy
solution is obtained from (\ref{solution0energy}) by taking $p'$ to
the horizon, or $\tau'\goesto\infty$, which can
then be thought as the effect of deforming the field at the
horizon. There is yet another way of saying this, which has a more
straightforward interpretation in terms of the BCFT, in which the
relation between $\tau$ and the boundary value of the tachyon field is
convoluted. Instead of placing $p'$ at the horizon, or $t=t'$ and
$\tau'\goesto\infty$, one can use the SL(2,R) symmetry to put $p$ at
$t=0,\,\tau=0$, and use the remnant symmetry to rotate
$\tau'\goesto\infty$ to $t_d'=i\pi\ell(n+\frac{1}{2})$, where $n$ is
an integer\footnote{One can see that $t_d$ is the image of the mapping
  since the value of $\mu$ between the two points is preserved.}. The
image of the mapping consists of all points since 
(\ref{solution0energy}) is periodic in imaginary time. One can then
think of the zero energy solutions as an array of fields deformations
at $t_d'$. We will have more to say about this interpretation below,
with a twist, when we consider the actual problem of the fermionic
field.
 
One should expect the anti-symmetric two point function of a
fermion field in a de Sitter invariant background to allow for a similar
decomposition. Because of the effects of the parallel transport
on spinors, however, one expects a dependence on
${\Psi_{\alpha}}^{\beta'}(p,p')$, which is the bispinor that
implements the parallel transport from $p$ to $p'$. We will take some
time to study its geometry. Following \cite{Allen:wd} and
\cite{Allen:qj}, we point out that $\Psi(p,p')$ is invariant under the
isometries of de Sitter -- what is called a {\it   maximally symmetric
  bispinor}. Then its derivative is also a maximally symmetric object
allowing for an expansion in terms of maximally invariant quantities:
\be
\nabla_{a}{\Psi_{\alpha'}}^{\beta}(p',p)=A(\mu)n_{a}(p',p)
      {\Psi_{\alpha'}}^{\beta}(p',p)+B(\mu)
      {{(\gamma_{c})}_{\alpha}}^{\beta}
      {{(\gamma_{a})}_{\delta}}^{\alpha}
      {\Psi_{\alpha'}}^{\delta}(p,p)
      n^{c}(p',p) \label{covariant1},
\ee
where $n_{a}(p',p)=\nabla_{a}\mu(p',p)$ is the normalized vector
tangent to the geodesic that links $p$ to $p'$. Primed indices live in
the ((co)spinor, (co)tangent) bundle at $p'$. We will be omitting the
spinor indices from now on, with the understanding that covariant
spinor indices live on $p'$. The bispinor $\Psi(p',p)$ is covariantly
constant over geodesics:
\be
n^{a}(p',p)\nabla_{a} \Psi(p',p)=0.
\ee
This means that $A(\mu)=-B(\mu)$ then (\ref{covariant1}) can be
rewritten as
\be
\nabla_{a} \Psi(p',p) = A(\mu)(g_{ab}-{\bf \gamma}_{a}{\bf
  \gamma}_{b})n^{b}(p',p) \Psi(p',p) 
=\frac{1}{2}A(\mu)[{\bf \gamma}_{a},{\bf \gamma}_{b}]n^{b}(p',p)
\Psi(p',p), 
\ee
or alternatively
\be
\Dslash\Psi(p',p)= A(\mu)\nslash\Psi(p',p).
\ee
In two dimensions one can further simplify the expression by
considering that the commutator of the gamma matrices should be
proportional to flat space ${\bf \gamma}^5$ and the volume element of the
space:
\be
\nabla_{a} \Psi(p',p)= A(\mu)\epsilon_{ab}n^{b}(p',p) {\bf
  \gamma}^5\Psi(p',p) 
\label{covariant2}.
\ee
One then needs to find the function $A(\mu)$. A condition for it comes
from the definition of curvature:
\be
[\nabla_{a},\nabla_{b}]\Psi(p',p)=\frac{1}{4}R_{abcd}{\bf
  \gamma}^{c} {\bf \gamma}^{d} \Psi(p',p) = \frac{1}{4\ell^2}[{\bf
    \gamma}_{a},{\bf  \gamma}_{b}]\Psi(p',p) ,
\ee
where we specialized to maximally symmetric spaces in the last
equality. Applying it to $\Psi(p',p)$ and using (\ref{covariant2}) we find:
\be
\epsilon^{ab}\nabla_{a}\nabla_{b}
\Psi(p',p)=\left(\frac{dA(\mu)}{d\mu} n^{a}n_{a} + A(\mu)\nabla_{a}
n^{a}\right){\bf \gamma}^5\Psi(p',p)=-\frac{1}{2\ell^2}{\bf \gamma}^5 
\Psi(p',p),
\ee
%remember Lorentz=\epsilon^{ab}\epsilon_{ab}=-2!
which, for timelike separations $n^an_a=+1$ yields a ordinary differential
equation for $A(\mu)$:
\be
\frac{dA(\mu)}{d\mu}+A(\mu)\nabla^2\mu=-\frac{1}{2\ell^2}.
\ee
By using the expression for the Laplacian in (\ref{laplacian}) one can
find $\nabla^2\mu$, and the equation has one single solution that
vanishes as $\mu\goesto 0$:
\be
A(\mu)=-\frac{1}{2\ell}\tanh\frac{\mu}{2\ell}.
\ee
One could in principle integrate $\Psi(p',p)$ using the
differential equation above with the boundary condition that
$\lim_{p\goesto p'}\Psi(p',p)={\bf 1}$ but this will not be necessary
for our purposes. 

The interpretation of the parallel transport operator is clear when
one considers the remarks of Section 3. The spectrum of a field in
de Sitter is divided into ``real'' excitations whose energy is larger
than the mass of the field, and ``virtual'' excitations, with energy
smaller than $m$. The former correspond to particles that actually
have enough energy to climb up the ``tachyon wall'' (\ref{potential}),
whereas the latter never actually make it, passing to some turn around
point at $p'$, its energy being redshifted by the non-triviality of
the metric, like the geodesic $B$ in Figure 1. These solutions where
labeled ``hyperbolic'' and ``elliptic'' respectively in
\cite{Larsen:2002wc}. The advantage of using the SL(2,R) symmetry as a
guiding principle is that one can relate the ``virtual'' spectrum at,
say the open string vacuum $\tau=0$ to the ``real'' spectrum at some
other point $p'$. The operator that implements this relation is $\Psi$. 

After this rather long detour one is able to deal with the equation
for the Hadamard function:
\be
(i\Dslash-m)\langle
   [\psi_{\alpha}(p),\bar{\psi}^{\beta'}(p')]\rangle = 
(i\Dslash-m) {S_{\alpha}}^{\beta'}(p,p')=0 .
\ee
Note that now the derivative acts on the covariant spinor index. By
writing  $S=-(i\Dslash +m)G$ and
${G_{\alpha}}^{\beta'}(p,p')=F(\mu){\Psi_{\alpha}}^{\beta'} 
  (p,p')$ 
one is able to get an ordinary differential equation for
$F(\mu)$: 
\be
\left(\Dslash\Dslash +m^2\right)F(\mu)\Psi(p,p')=
(\nabla^2F+(m^2-A^2(\mu)+\frac{1}{2\ell^2})F)\Psi(p,p')=0 .
\ee
The equation itself is:
\be
\left[\frac{d^2}{d\mu^2}+\frac{1}{\ell}
  \frac{\cosh\frac{\mu}{\ell}}{\sinh\frac{\mu}{\ell}}
\frac{d}{d\mu}+\left(m^2-\frac{1}{4\ell^2}\tanh^2\frac{\mu}{2\ell}
+\frac{1}{2\ell^2} \right)\right]F(\mu)=0,\label{equationH}
\ee
whose solutions are another set of hypergeometric
functions. We present the one suitable for the Hadamard
elementary function, which has a rather lengthy expression for
$\mu^2>0$:
\begin{eqnarray}
F(\mu)&=& -\frac{1}{8\pi}\cosh\frac{\mu}{2\ell}
\left[{_2F_1}(1+im\ell,1-im\ell;1; 
-\sinh^2\frac{\mu}{2\ell})\log\left(\sinh^2\frac{\mu}{2\ell}\right)+
\right. \nonumber \\
& & \quadsp\quadsp
\left.\sum_{n=0}^\infty\frac{(1+im\ell)_n(1-im\ell)_n}{(n!)^2} 
K(n,m\ell)(-1)^n\sinh^{2n}\frac{\mu}{2\ell}
\right]\label{hadamard}\!\!,
\end{eqnarray}
where $(z)_n=\Gamma(z+n)/\Gamma(z)$ is the Pochhammer symbol, and
$K(n,m\ell)= \psi(n+1+im\ell)+\psi(n+1-im\ell)-2\psi(n+1)$,
$\psi(x)=\frac{d}{dx}\log \Gamma(x)$ being the digamma function. 
The constant of normalization being fixed by
requiring that the function has the flat space form near the wave
front $\mu=0$.
\be
F(\mu)=-\frac{1}{8\pi}\log \mu^2 + {\cal O}(\mu^0). \label{dw-s-exp}
\ee
For spacelike separations, one is helped by the fact that the
monodromy of the solutions around the singular point $\mu=0$ is
parabolic. The relative sign coming from the logarithm term cancels
the one coming from the hyperbolic sine. One then can use well-known
properties of hypergeometric functions to write $F(\mu)$ in a more
compact form:
\be
F(\mu)=\frac{1}{8\pi}\frac{\pi m\ell}{\sinh \pi m\ell}
\cosh\frac{\mu}{2\ell}\,{_2F_1}(1+im\ell,
1-im\ell;2;\cosh\frac{\mu}{2\ell}),\quadsp\mu^2<0
\ee
from which we observe for future reference that $F(\mu)$ is regular
when $\mu=in\pi\ell$. 
 
Since the argument of the hypergeometric can be written as
a function of $Z$, one sees that (\ref{hadamard}) has poles when
$Z=1$. The ``vacuum'' of the tachyon field has the (anti) periodicity
in imaginary time given by the periodicity of $Z(t,\tau)$, which in
turn is $t\goesto t+iT_H^{-1}$, with:  
\be
T_H = \frac{1}{2\pi\ell}\label{temperature},
\ee
which is generically assigned to thermal backgrounds. We will however
take another interpretation of this fact. The important quantity from
the SL(2,R) point of view is not $t$ but $\mu$. And inspecting the
function above one sees that it is invariant under $\mu\goesto
\mu+2\pi i\ell$, that is to say that space-like separations are {\it
  compactified} with period $T_H$. This means that we can take the
global structure of this de Sitter space-time to be that of a
cylinder, like (\ref{cylinder}) with $\epsilon=1$. From
(\ref{hadamard}) one sees that the fermion field is anti-periodic in 
this direction. 

Consider the zero energy solutions{\footnote{In usual BCFT these have $\lambda=\frac{1}{2}$.}} 
of the quantum-mechanical system: 
\be
\chi^{\pm}_0= e^{-2m\ell\, {\rm arctan}\, e^{\pm
    \ell^{-1}\tau}}\eta_{\pm}. 
\ee
These solutions are not normalizable in the equivalent quantum mechanics
problem (\ref{susyqm}), which means that their degrees of freedom are not
dynamical, but their action (\ref{clasvalueaction}) is finite:
\be
S_{\rm cl}(\chi_0)= i\pi m\ell .
\ee
Each wave function is a superposition of chiral
components of each side of the tachyon wall and are supported at
$\tau=\pm\infty$, having exponential tails at the other side. We will
thus identify each of them with half-(anti)-D-instantons. Zero
``global'' energy solutions are removed from the cylinder by the
anti-periodicity condition but we can still consider it as the limit
where one takes the bounce solution to the horizon, where its energy
is redshifted away. The creation (annihilation) operators are
interpreted as half-(anti)-D-instantons. Since we argued above that
the two-point function is periodic in space-like distances, placing
these field insertions at the horizon means placing them at $\mu=2\pi
in\ell$, $n$ being any integer. The insertion of a pair of instantons
and anti-instantons as described above will give rise to a sD-brane.  
These correspond to the ``imaginary D-branes'' background as
considered in \cite{Gaiotto:2003rm}, which indeed were found to give
rise to purely closed string amplitudes in BCFT formalism. In the
language above this is natural since the insertions create field
quanta at the horizon.

The solution (\ref{hadamard}) above was derived under the assumption 
that the singularity happened when $\mu=0$, or $Z=1$, and not
$\mu=i\pi\ell$, or $Z=-1$, in which case one allows for the second
solution of (\ref{equationH}):
\be
\tilde{F}(\mu)=\tilde{N}\cosh\frac{\mu}{2\ell}\, {_2F_1}(1+im\ell,1-im\ell;1;
-\sinh^2\frac{\mu}{2\ell}).
\ee
in de Sitter the solutions with poles at
$\mu=in\pi\ell$  are normalizable and extra assumptions (like a
Hadamard development of the pole at $\mu=0$) are required to drop
it. This general case corresponds to a fermionic analogue of the
``alpha vacua''.

So the question that naturally arises is: just what these poles at
$Z=-1$ correspond to in the usual boundary state formalism? The extra
pole of the alpha vacuum can be seen as putting a particle in a state
which is a superposition between a given point $p$ and its antipodal
point $p^A$, which is defined through the embedding of de Sitter in
the higher dimensional Minkowski space-time with coordinates
$\vec{z}=\{z_0, z_1, z_2\}$
\cite{Allen:ux}:
\be
z_0=\ell\frac{\sinh \ell^{-1}t}{\cosh \ell^{-1}\tau},\quadsp
z_1=\ell\frac{\cosh \ell^{-1}t}{\cosh \ell^{-1}\tau},\quadsp
z_2=\ell\frac{\sinh \ell^{-1}\tau}{\cosh \ell^{-1}\tau}. 
\ee
The de Sitter space is then the hyperboloid
$z^2_0-z_1^2-z_2^2=-\ell^2$. Generically, the antipodal point
$\vec{z}^A=-\vec{z}$ lies outside the coordinate patch covered by
$t,\tau$.  For the excitations we are interested on, however, are
localized near $\tau=\infty$, the image of the map is localized near
$\tau=-\infty$, and time-reversed. So in these cases the fermion
density will be non-zero for both sides of the potential. From the
discussion above and the work of \cite{Gaiotto:2003rm} the generic
SL(2,R) invariant state seems to be related to a generic configuration
of (anti)-D-instantons, giving rise to generic background values for
the closed string tachyon and the RR boson. It would be interesting to
provide a clear relation between these states and backgrounds in the
closed string formulation.

Quantum effects raise the temperature of the vacuum by $T_H$
(\ref{temperature}), an effect which should be, as discussed in the
Section 3, independent of the coordinates used to parame\-trize the
D-branes trajectories in phase space. As the conjecture goes in
\cite{Douglas:2003up}, the matrix model at finite temperature
corresponds to a different closed string background where 
the Liouville interaction is perturbed by $\Lambda e^{\varphi}$ and not by
$\varphi e^{\varphi}$ \cite{Klebanov:1994kv}. We will have more to say
about this point below. This temperature (\ref{temperature}) actually
corresponds, for type 0B, to the self-dual point on which vortices in
the finite matrix model coexist with the singlet configurations
discussed here \cite{Gross:1990ub}. Further understanding of the
bosonization of the system (\ref{effaction}) could provide another
view of this effect.

One can also use the two-point function to compute the
backreaction. The expectation value of the stress-energy $\langle
T_{ab}\rangle$ requires the regularization of the two-point 
function for coincident points. This is a standard calculation in QFT
in curved space \cite{Candelas:du}. Firstly, since the state is
invariant under SL(2,R), the expectation value of the stress tensor
will be proportional to the metric. Then one only needs to compute the 
expectation value of the trace of the stress tensor:
\be
\langle g^{ab}T_{ab}\rangle =\frac{i}{2}\lim_{p\goesto p'}
\langle\psi_{\alpha}(p){(\gamma^{a'})_{\beta'}}^{\alpha}\nabla_{a'}
\bar\psi^{\beta'}(p')
-\nabla_{a}\psi_{\alpha}(p){(\gamma^{a})_{\beta'}}^{\alpha}
\bar{\psi}^{\beta'}(p')\rangle,
\ee
which can be rewritten in terms of $S$:
\be
\langle g^{ab}T_{ab}\rangle_{\rm ren}=\frac{1}{2}\lim_{p\goesto p'}
{\rm Tr} (i{\bf \gamma}^a\nabla_a S_{\rm ren} - i\nabla_{a'}S_{\rm
  ren}{\bf \gamma}^{a'}) \label{trace},
\ee
where one regularizes the two point function in the usual way:
subtracting an infinite piece of it. This is usually accomplished in
QFT in curved space-time by redefining the couplings in the full
Lagrangian with the Einstein-Hilbert term. The practical implication
of this is that one subtracts from the Hadamard function a purely
``geometrical'' contribution, usually called the DeWitt--Schwinger, or
adiabatic, expansion. Here we are not really allowed to do this since
gravity is non-dynamical in two dimensions. The ultra-violet
divergence will at any rate generate a term in the renormalized
Lagrangian, whose form we will now argue to be a Liouville perturbation
term. This also means that only the diverging term is of immediate
physical interest to us, and we will not discuss the finite
pieces. Expanding (\ref{hadamard}) and inserting it into (\ref{trace}),
we will have:
\be
\langle g^{ab}T_{ab}\rangle_{\rm div} = 2\lim_{\mu\goesto 0^+}
\frac{m^2}{8\pi}\log\frac{\mu^2}{4\ell^2}+ \mbox{finite terms}.
\ee
The factor of $2$ comes from the trace over spinor indices. The
appearance of the Liouville term can be understood heuristically as
follows: in the language of \cite{birrell}, it ``renormalizes'' the
cosmological constant and the Einstein term in $2+\epsilon$
dimensions. One can then arrive at the usual form of the Liouville
Lagrangian upon dimensional regularization, as in
\cite{DaCunha:2003fm}. The procedure does not quite work here,
however, since the Lioville field has the usual interpretation as the
field obtained by bosonization of the Lagrangian (\ref{effaction}),
but the form of the perturbation should still be the same, as one
could argue from the fact that the system, being a CFT, must have a
well-defined UV completion. The appearance of Liouville-like terms is
universal in two dimensional systems \cite{Ginsparg:is}, and here is 
an example which actually bears resemblance to the massive Ising
model. Since the redefinition of the density of eigenvalues as
profiles of the Liouville field involves a $m$ dependent term, as well
as an exponential coordinate transformation, the first term can be
taken as inducing the generation of a perturbation like $m e^\varphi$
in the effective bosonic model. 
%Starting from the divergent piece of the Feynman function as in
%\cite{Candelas:du}, we arrive through dimensional regularization to:
%\be
%8\pi\sqrt{-g}\frac{\partial {\cal L}_{\rm div}}{\partial m^2} =
%-\frac{1}{\epsilon}
%-\psi(1)+1-\log\frac{4\pi}{\ell^2}+\psi(im\ell)+\psi(-im\ell) + {\cal
%  O}(\epsilon),
%\ee
%where the dimension of the space-time is $n=2(1-\epsilon)$. 

\section{Qualitative remarks}

We saw that even though the DBI action is just an effective theory to
the dynamics of D-branes in a tachyon background, one can still infer
a lot about closed string dynamics from it. The reinterpretation of
the DBI action as describing particles moving in a two-dimensional de
Sitter space-time allows for both qualitative and quantitative
tests in the relation of the double-scaled matrix model and type 0B in
two dimensions. The most enticing prospect for future work is the
bosonization of (\ref{effaction}), which will clarify the relation
between the DBI and closed string dynamics and to Liouville. It would
also be interesting to use mode sum techniques to compute the
effective action in the ``Rindler'' state found in Section 4. This
will provide a means to compute the backreaction due to the
condensation process, with the additional nice feature that the string
coupling can be made finite with no extra complications. Yet another
route one could take would be to interpret the thermal two-point
function found above as the result of insertions of closed string
(Liouville) boundary operators. This would give a DOZZ
picture of the sD-brane state and would help distinguish the different
vacua in the model of de Sitter quantum gravity provided by Martinec 
\cite{Martinec:2003ka}. 

The point is made that tools used in the
studies of QFT in curved space-time can help developing the
McGreevy-Verlinde interpretation of the tachyon condensation
process. In fact, one has generically that the tachyon potential
decays exponentially near the closed string vacuum. By the point of
view taken in this paper, this amounts to considering a metric of the
form
\be
d\hat{s}_{\mbox{\scriptsize eff}}^2=V^2(T)(-dt^2+dT^2)= e^{-2a
  T}(-dt^2+dT^2)
\ee
where $a>0$. This corresponds to Rindler space, which lends itself to
the same type of treatment given here: there is an infinite
backreaction of the Rindler vacuum, and there is a analogue of an
invariant state which is seen by the Rindler observer,{\it i.e.}, the
one following curves of constant $T$, as a thermal bath of
radiation. In fact in this case the ``invariant state'' is none other
than the usual Minkowski vacuum. The model above should be equivalent
to models of relativistic fermions considered before, as in
\cite{Klebanov:2003km}, but somehow the importance of the Rindler
state has been so far overlooked.

It is curious in this case to recover a non-flat background without
what one would na\"ively call gravitational degrees of freedom. The form
of the potential in this case comes directly from the string field
theory, and in this sense the de Sitter background can be considered
``on shell''. ``Off shell'' configurations $\Psi_o$ would arise from the
partition sum of the string field theory in the usual fashion, {\it
  i.e.} having a Boltzmann weight given by $e^{-S(\Psi_o)}$, with the
open string field theory action. The novelty here is that this
formulation does not seem to lend itself to a straightforward
interpretation of fluctuations as gravitons, but one expects a
Liouville-type of theory based on diffeomorphism invariance. Even
though the form of the tachyon potential gives rise to conformally
flat backgrounds, this interpretation only gives a meaningful
prescription in two dimensions, and even then there are global issues
in field space which are far from clear. We have not considered the
effect of non-trivial RR backgrounds, either. 

At any rate, one cannot help but wonder at the coincidence in
terminology between this interpretation of the tachyon decay process
and the usual open-closed string duality, as in the ADS-CFT. There,
the near-horizon limit of the solution of the effective closed string
theory (SUGRA) equations gave an ``on-shell'' background which could
alternatively be described as open strings degrees of freedom. Here,
the effective {\it open string} degrees of freedom could be seen as if
propagating in a non-flat background, de Sitter in the present
case. The curved background arises as an ``on-shell'' configuration as
far as the string field theory equations are concerned. The ``near
horizon'' limit here appears as one considers excitations of the
branes with low energy, which would translate to excitations near
$\tau=\pm\infty$ in the present interpretation. The ``near horizon''
limit then reveals {\it  closed string} degrees of freedom, in a
construct that resembles recent holographic formulations in de
Sitter space \cite{Banks:2003kr}, \cite{Parikh:2004blah}. Remarks on
this regard have been made by Sen in \cite{Sen:2003xs}. There seems to
be a lot of important physics one can extract from the DBI action with
relation to both QFT in curved space-time and the fate of the tachyon
condensation in two dimensions. One hopes that this will help
shed light on the generic problem.

\section*{Acknowledgments}

I would like to thank Shinsuke Kawai, Esko Keski-Vakkuri, Archil
Kobakhidze, Juan Maldacena, Emil Martinec and Martin Sloth for comments and
suggestions.

\section*{Appendix: Conventions on Spinor Calculus}
We will follow gamma-matrix based approach. In a manifold $M$ one
could define a spinor bundle under some topological conditions. For
each element of the spinor bundle $\psi^\alpha\in S$ and another element of
the dual spinor bundle one $\phi_\beta$ can define an element of the
(co)tangent bundle $v_a$ by means of the gamma matrix. The gamma
matrix itself is then a 1-form valued in the tensor product $S\times
S^*$\footnote{This in turn is equivalent to the space of linear
  operators in $S$. We will be taking this equivalence to be trivial.}:
\be
v_a=\phi_\beta{{(\gamma_a)}_\alpha}^\beta \psi^\alpha,
\ee
which will be taken to satisfy the algebra (omitting spinor indices):
\be
\{{\bf \gamma}_a,{\bf \gamma}_b\}=2g_{ab}{\bf 1},
\ee
where $g_{ab}$ is the metric of the manifold and ${\bf 1}$ the identity on
the spinor bundle. The gamma matrix one-form can be written in terms
of the vielbein $e^i_a$ and the flat space gamma matrices ${\bf \gamma}_i$
as:
\be
{\bf \gamma}_a = \sum_i e^i_a{\bf \gamma}_i.
\ee 
The gamma matrix one form also defines an affine connection on spinors
by the imposing that the Clifford algebra structure is covariantly
constant: 
\be
\nabla_a{{(\gamma_b)}_\alpha}^\beta= \bar{\nabla}_a
      {{(\gamma_b)}_\alpha}^\beta - C_{ab}^c
      {{(\gamma_c)}_\alpha}^\beta - {{(\Gamma_a)}_\alpha}^\delta
      {{(\gamma_b)}_\delta}^\beta + {{(\gamma_b)}_\alpha}^\delta
      {{(\Gamma_a)}_\delta}^\beta = 0 \label{connection}.
\ee
By relating the quantities with $\bar{\nabla}_a$ being the flat space
derivative, one arrives at the well-known formulas for the Christoffel
symbols and the spinor connection:
\be
{\bf \Gamma}_a = \frac{1}{4} \omega_{aij}{\bf \gamma}^i{\bf \gamma}^j,
\ee
with $\omega_{aij}$ being the spin connection, defined from the
vielbien by the zero torsion condition:
\be
de^i_a+{\omega_{aj}}^i\wedge e^j_b=0,
\ee 
so ${\omega_{aj}}^i=e^b_j\nabla_a e_{bi}$. As in usual tensor
calculus, the commutator of two derivatives of a spinor satisfies
linearity:
\be
[\nabla_a,\nabla_b](A(x)\psi(x)+B(x)\eta(x))=
A(x)[\nabla_a,\nabla_b]\psi(x) +B(x)[\nabla_a,\nabla_b]\eta(x),
\ee
and in fact can be computed in terms of the spin connection:
\be
[\nabla_a,\nabla_b]\psi = (\partial_a {\bf \Gamma}_b-\partial_b{\bf
  \Gamma}_a - {\bf \Gamma}_a{\bf \Gamma}_b + {\bf \Gamma}_b{\bf
  \Gamma}_a)\psi = S_{ab}\psi .
\ee
And by considering that the one-form
$\omega_a=\xi_\beta{{(\gamma_a)}_\alpha}^\beta\psi^\alpha$
satisifies:
\be
[\nabla_a,\nabla_b]\omega_c={R_{abc}}^d\omega_d,
\ee
one can relate the tensors $S_{ab}$ and $R_{abcd}$ by using the chain
rule and some properties of the gamma matrices:
\be
S_{ab}=\frac{1}{4}R_{abcd}{\bf \gamma}^c{\bf \gamma}^d.
\ee
The action on covariant spinors can be obtained similarly. For instance:
\be
[\nabla_a,\nabla_b]\psi_\alpha=-\psi_\beta{(S_{ab})_\alpha}^\beta.
\ee
And we finish by listing properties of the parallel transport operator
${\Psi_\alpha}^{\beta'}(p,p')$. We will convention that $p$ and  $p'$
refer to the covariant and contravariant indices, respectively. In the
formulas below $\mu$ is the geodesic length between $p$ and $p'$ and
$n^a=\nabla^a\mu$ is the unit vector tangent to the geodesic.
\be
\lim_{p\goesto p'}\Psi(p,p')={\bf 1},
\ee
\be
\Psi(p,p´)\Psi(p',p)=-{\bf 1},
\ee
\be
n^a\nabla_a\Psi(p,p')=0,
\ee
\be
\nabla_{a'}\Psi(p,p')=\frac{1}{2}A(\mu)[{\bf \gamma}_{a'}, {\bf
    \gamma}_{b'}]n^{b'}\Psi(p,p'),\quadsp\quadsp A(\mu)=
-\frac{1}{2\ell}\tanh\frac{\mu}{2\ell},
\ee
\be
\nabla_{a}\Psi(p,p')=-\frac{1}{2}A(\mu)\Psi(p,p') [{\bf \gamma}_{a},
  {\bf \gamma}_{b}]n^{b}. 
\ee
We also use the ``slash'' notation. For the sake of completeness,
$\Dslash \psi={\bf \gamma}^a\nabla_a\psi$ and $\Dslash
\psi^T=\nabla_a\psi^T{\bf \gamma}^a$ for contravariant and covariant
spinors, respectively. For a more detailed consideration of
$\Psi(p,p')$ see \cite{Allen:qj}.

\end{document}